# Soft Computing – A step towards building Secure Cognitive WLAN

S.C.Lingareddy[1]
Asst. professor and HOD, CSE,
KNS Institute of Technology,
Bangalore, India.

Dr B Stephen Charles[2]
Principal
Stanley Stephen College of Engg,
Kurnool, India.

Dr Vinaya Babu[3]
Professor of CSE,
Director of Admission Dept.
JNTU, Hyderabad, India.

Kashyap Dhruve[4]
Technical Director
Planet-i Technologies
Bangalore, India.

*Abstract*— Wireless Networks rendering varied services has not only become the order of the day but the demand of a large pool of customers as well. Thus, security of wireless networks has become a very essential design criterion. This paper describes our research work focused towards creating secure cognitive wireless local area networks using soft computing approaches. The present dense Wireless Local Area Networks (WLAN) pose a huge threat to network integrity and are vulnerable to attacks.

In this paper we propose a secure Cognitive Framework Architecture (CFA). The Cognitive Security Manager (CSM) is the heart of CFA. The CSM incorporates access control using Physical Architecture Description Layer (PADL) and analyzes the operational matrices of the terminals using multi layer neural networks, acting accordingly to identify authorized access and unauthorized usage patterns.

I. INTRODUCTION

Wireless Networks have become an integral part of the Information Technology (IT) Infrastructure especially Wireless Local Area Network (WLAN). The Institute of Electrical and Electronics Engineers (IEEE) have described certain standards and specification for efficient communication over the wireless medium IEEE 802.11[26]. WLAN's have seen a tremendous growth because of their ability to provide flexible and mobility options to the user. WLAN's are now an integral part of both Enterprise Network and Public Networks. With such dense WLAN's available to user's, security is indeed a major concern. The security features provided in the current deployed WLAN's are vulnerable [1].

Cognitive Networks also known as smart networks [2] could be a solution to this security and data integrity concern. Cognitive Networks are recognized for their self aware, self management and self healing properties. Wireless networks have been studied as complex, heterogeneous and dynamic environments and cognition of these are still under research [3]. Cognitive networks could be used to improve resource management [4],[17], quality of service (QoS)[5], security and access control[6]. We propose to introduce a CFA to secure WLAN's.

Computational Intelligence is defined as the study and design of intelligent components. To impart computational intelligence we propose to use soft computing techniques. Soft Computing encapsulates various several intelligence imparting technologies including fuzzy logic, neural networks, probabilistic computing, artificial immune systems etc. We intend to use multilayer neural networks in our CFA.

In this paper we put forward the CFA where we would use multiple layers of neural networks to impart intelligence to the proposed framework. The intelligence parameters of the CFA are analyzed in the Experimental Setup.

II. RELATED WORK

WLANS are one of the most lucrative and fast growing deployments for connectivity options for networks. The WLANS are established by Access Points (AP). The AP that are provided by the manufactures or service providers. Based on the literature provided to the users they feel that the AP's are secure which is not the fact [7]. The AP's provide security in multiple ways like WEP [8], [22], WPA [9] ,TKIP[10], etc which are vulnerable and could be easily rendered ineffective. Even the authentication schemes provided in the AP, based on the user node hardware are ineffective. The current AP's provide Ethernet MAC based filtering for access control but this too was found to be insecure [11]. It is very clear that the WLAN has been known for its non secure nature [7]. Many protocols have been developed with a hope to provide better security [27]. However, much research is ongoing to solve this security issue [10][13][14].

Based on our research we believe that these WLANS could be made intelligent and the access control mechanism could be improved to negate these security deficiencies. Much work has been done towards securing wireless networks based on cognitive approaches. When we speak of cognition terms like software defined radio and cognitive radio [4], they are often misunderstood. Software defined radio is simply the radio layer which transmits the radio frequencies and intermediate frequencies. Cognitive Radio on the other hand rests above the software defined radio layer and is intelligent. The cognitive radio layer controls the software defined radio and determines which modes of operation to be assigned to the software defined radio.

Cognitive Networks are different from cognitive radio. Based on the research carried out we have studied that cognitive networks are based on cross layer optimizations that have been successful as they alter the parameters of multiple layers of the protocol stack. Efficient work has been carried out on the inter layer or cross layer protocols [16]. A lot of work is done towards providing security based on cognitive radio[17][18][19]. Many researchers have proposed construction of cognitive networks using cognitive agents[20]. Cognitive agents are integrated to provide intelligence to the networks.

Cognitive Networks design is an area of tremendous interest. Network design based on the Observe Orient Decide Act loop [2] gives us a clear understanding of the CP. Node behavior analysis [20], Node reputation [15] is also an eminent criterion taken into consideration for providing security and network cognition. In our approach we recommend to use lightweight and cooperative algorithms which have cross layer optimization and can be considered towards structuring cognitive networks [14] [3]. Access control is a very important aspect to be considered with respect to security and network integrity [6].Access control in many approaches has been provided by additional servers which might be a tradeoff between security and quick network access [24]. Network Integrity could be maintained very efficiently provided strong access control mechanisms are incorporated to the network design. We propose an access control mechanism based on PADL [21] based on our previous research.

From our extensive research is it very evident that soft computing approaches have been incorporated to provide human intelligence into networks [25]. Soft Computing is a wide area of research today which includes neural networks, fuzzy logic artificial immune systems etc. Learning and reasoning are successfully implemented using neural networks. Studies have shown that many systems designed using neural networks have proved robust and effectively handle the CP.[17][5][21][23]. Neural Network design and its incorporation into the cognitive networks are considered as the most challenging aspects of the CP [28]. Neural Networks training algorithms could be classified into two categories - supervised and unsupervised learning. In our scheme we intend to use multi scalar neural networks. Fuzzy Logic, Genetic Algorithms [17], Game Theory, Markov Chain, pricing theory based approaches have also been studied.[14].We also propose to use a multi layer neural network trained using the back propagation algorithm.[23]

### III. PROPOSED SYSTEM

The security issues with the current implementation of the WLAN have clearly been stated earlier. Cognitive Networks can be a solution to the current issues of WLAN's. Cognitive Networks could improve resource management, QoS, security and access control. Here we would discuss about a CFA we propose, where our motive was to maintain a controlled network framework [6].The CSM which is the heart of the CFA is designed to maintain the cognitive network and also implements the CP using neural networks. A very important factor that was considered is that a network should understand what the application can achieve and an application should be capable of understanding what the network is capable of doing at any given point in time. Basically for successful construction of a CSM joint layer, optimization and cross layer adaptive design is the key factor of consideration. A controlled and monitored environment is created to maintain the reliability of the network.

The major factors considered for the CSM proposed are towards secure access control mechanisms and implementation of the CP. The proposed CFA for WLAN's is as shown in fig.1.

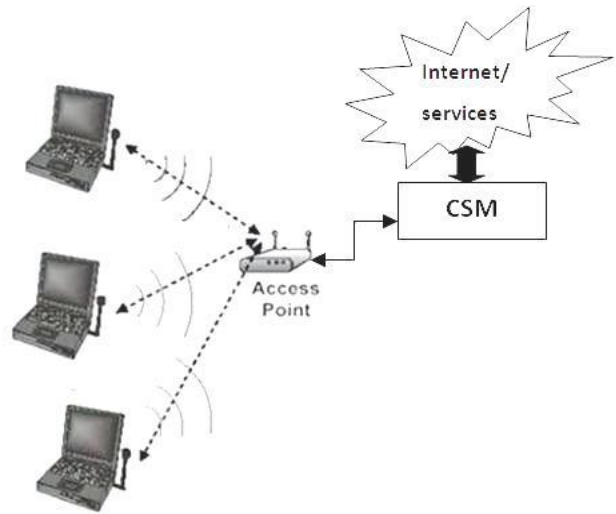

Figure 1: Cognitive Framework Architecture

From Fig.1 it is clear that the CSM imparts cognition to the network. The CSM implements a controlled and monitored WLAN access. A cross layer adaptive design is used to implement the CSM. The Block Diagram of the CSM is as shown in Fig.2.

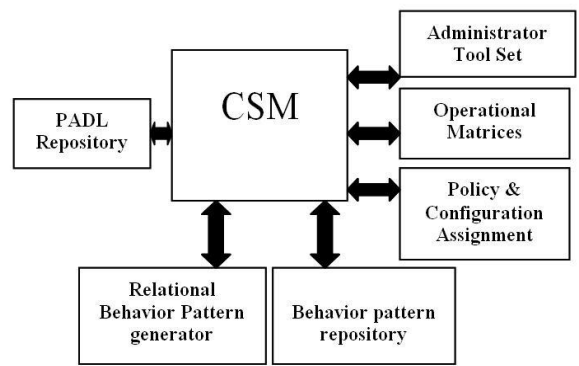

Figure 2: CSM Architecture

Access Control is the key parameter considered to maintained network integrity and controlled network access. Attacks on AP's, node jamming, node misbehavior etc are known problems for cognitive networks. These problems could be negated by providing strong authentication to its users and also by having complete knowledge of the network components. The CSM provides strong access control mechanisms for the CP. The CSM is aware of the network distribution architecture. The CSM is also well equipped with strong user authentication schemes. But the CSM is unaware of the terminals and its user authenticity.

The CSM identifies the terminals based on its Physical Architecture Description Layer (PADL). The PADL is a collection of data from the physical layer and the radio layer of the terminals as shown in our previous research [21]. The PADL would be unique for each terminal, which would enable the CSM to identify the terminal for Node misbehavior, unauthorized access, jamming etc. The CSM through the PADL could maintain complete knowledge of the network components and construct a secure controlled environment for network transactions.

The CSM maintains a PADL repository. The PADL repository houses two sections. One section houses the PADL of authorized terminals and the other section for unauthorized terminals. If an authorized terminal's user misbehaves or his actions pose a threat to network integrity, the PADL of the node is moved to the unauthorized section of the PADL repository. The user misbehavior is then detected by the Policy Manager of the CSM. Here the importance of joint layer optimization could be clearly understood. The CSM is also responsible for network management, monitoring, analyzing the user patterns and imparting the administrative user matrices.

The administrative tool set is controlled by the network administrator. Nodes in the CFA are recognized by the CSM by their PADL. Based on the PADL they are classified into authorized nodes, unauthorized nodes and new nodes. New nodes are nodes whose PADL is not found in the authorized and unauthorized section of the PADL repository. When a new node arrives the CSM interacts with the Administrative Tool Set to initialize this new node into either an authorized node status or unauthorized node status, based on the administrator's discretion. The administrative tool set also initializes the operational matrix of the new registered node using a very conservative approach protecting the network resources and the services it offers. The operational matrices generated are stored in the operational matrix repository.

The administrative tool set has another function. The network is maintained by administrators who can at any point alter the configuration and the operation of the CSM using this administrative tool set.

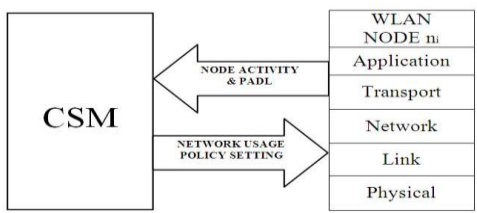

Figure 3: Cross Layer CSM Adaptive Design

An important factor taken into consideration for maintaining is that the CSM has to gather and retain information of the network activities continually, to effectively realize the CP. The cross layer adaptive design could be easily understood by Fig.3.

The CSM receives the node activity and provides the activity details to the Relational Behavior Pattern Generator. It generates a behavior pattern using a multilayer neural network. A simple neural network is as shown in Fig.4.

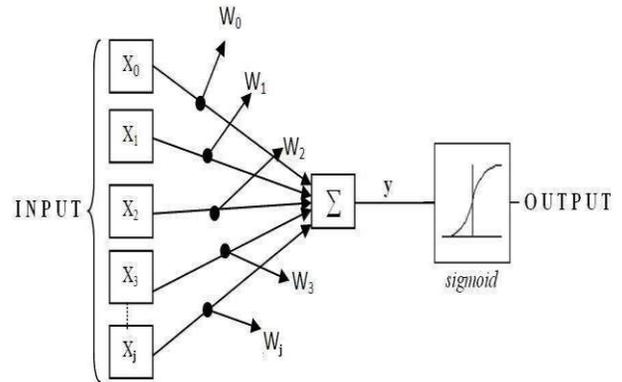

Figure 4: Neural Network with Sigmoid

$$y = \sum_{i=0}^{j} x_i w_i \qquad (1)$$

Equation (1) represents the output of the neural network shown in Figure 4. Y is the output of the neural network. $X_i$ is the input vector provided by the CSM using the node activity. J denotes the number of hidden neurons. The weights obtained from the operational matrix repository are represented as $W_i$. The neural network is trained using the back prorogation algorithm. Each registered node has an independent operational matrix. Equation (2) represents the sigmoid function also known as the activation functions. The output pattern given is Equation (2) is then taken by the CSM for analysis.

$$OUTPUT = \frac{1}{1 + e^{-y}} \qquad (2)$$

The CSM maintains the behavioral usage patterns of every node in the Behavior Pattern Repository. The pattern repository contains the previous behavioral patterns of the nodes this is used as the training set for the Policy and Assignment Management Unit.

A Multilayer Feedforward Neural Network (MFNN) is used to analyze the current node behavioral pattern. MFNN's are a valuable mechanism to analyze real time communication data. The MFNN also exhibit effective learning capabilities which are very essential to achieve cognition. A simple MFNN is as shown in Fig.5

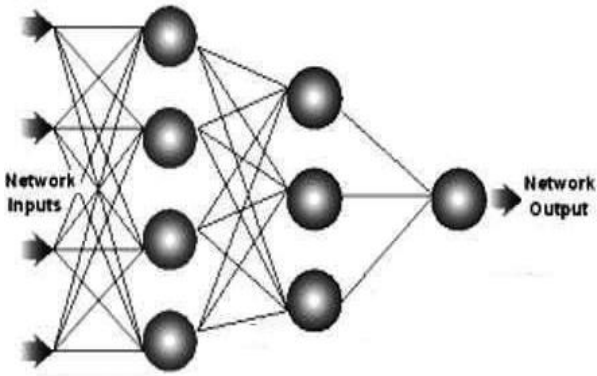

Figure 5. Multilayered Feedfoward Neural Network

The MFNN in the Policy and Configuration Assignment block of the CSM imparts the intelligence to the CP. The MFNN accepts the current node behavior pattern as an input and the previous operational usage patterns which are obtained from the Behavior Pattern Repository as the training set. The MFNN is trained using the back prorogation algorithm. The MFNN used has 2 hidden layers. The MFNN could be used to detect any misbehavior, any unauthorized network service access etc. The MFNN on detection of misbehavior changes the node status to unauthorized. The CSM on detecting an unauthorized node eliminates the node's PADL related data in the Behavior Pattern repository and Operational Matrices repository. Then the CSM moves the unauthorized PADL from the registered nodes to unregistered node within the PADL repository.

The algorithm developed for the CP is as given below. The processes that are executed when a terminal or node $N_i$, whose PADL is represented by $PADL_i$ and the node activity is represented as $NA_i$. The operational matrix if $N_i$ is $OM_i$. Node $N_i$ behavioral pattern is represented by $BH_i$ and its previous patterns used for training is represented as $TBH_i$.

Algorithm of CSM operation

1. Node $N_i$ enters the Wireless Network Environment
2. $PADL_i$ of the node is obtained by the CSM
3. The CSM classifies $PADL_i$ into 3 categories
   Case i $PADL_i$ = New Node
   Case ii $PADL_i$ = Authorized Node
   Case iii $PADL_i$ = Un-Authorized Node

4. **Switch (Case)**
   {

   **Case i:**

   a) CSM contacts the Administatrative Tool Set, which provides a secure or conservative operational matrix $OM_i$ to the CSM.

   b) CSM stores the $OM_i$ in the Operational Matrices Repository and also the $PADL_i$ in the registered Section of the PADL repository.
   **Break;**

   **Case ii:**

   a) The CSM obtains the operational Matrix ($OM_i$) from the operational matrices Repository based on $PADL_i$.

   b) The CSM obtains the node activity $NA_i$ from the Node $N_i$.

   c) The CSM provides the $NA_i$ & $OM_i$ to the Relational Behavior Pattern Generation.

   d) The Behavioral pattern BHi is obtained by the CSM

   e) The CSM based on the PADLi obtained from the training set of the Node Ni from the Behavioral Pattern Repository TBHi.

   f) The CSM sends TBHi and BHi to the Policy & Configuration Assignment block for neural analysis.

   g) The policy and Configuration assignment block analyzes $BH_i$ with the previous behavior history $TBH_i$.

   h) **If** (behavior variation is less than the threshold $\theta$ (set by the Administrative tool set)).
   {
       i. The Node Behavior is Normal
       ii. The CSM stores the BHi into the behavioral Pattern Repository
   }
   **Else**
   {
       i. The Node $N_i$'s network activity is deviated from its normal behavior.
       ii. Policy for Node status is changed to Unregistered
       iii. The CSM deletes the PADLi based entries in the Operational Matrices Repository and Behavioral Pattern Repository. The CSM moves the PADLi from the registered section to the Unregistered section.
   }
   **Break;**

   **CASE iii:**
   a) Network services and N/w resource usage to the node is prevented by the CSM.
   **Break;**
5. Go to 1

In our proposed system, the CFA provides strong access control mechanism, detection of node misbehavior techniques efficient for security mechanisms. The CP we proposed adopts a conservative approach and maintains the network security by establishing a controlled network. We evaluated the performance of the CSM in our experimental study.

## IV. PERFORMANCE EVALUATION

In order to illustrate the concepts discussed in the paper we have considered a simple wireless test bed similar to the configuration shown in Figure (1). We have recorded the Node Activity of 60 nodes in the WLAN. The WLAN access was distributed by using 6 access points. We had considered that the CSM offers access to data servers and also provides internet access services. The CSM records the node activity provided by the 60 Nodes. The Node activity taken into consideration was specific to the services distributed through the CSM. The CSM developed was deployed on a Quad-Core Server with 4GB of RAM.

We had considered similar Operational Matrices applicable to all the test nodes. The Relational Behavior Pattern Generator is developed using a multilayer neural network with the sigmoid activation function. Back Prorogation was used in training the network. To decide the configuration of the neural network we had to evaluate the performance of the neural network across various aspects such as the number of Neurons to be used in the input layer, the learning rate of the neural network and the number of iterations required. For efficient Behavioral pattern generation it is necessary that the neural network have a less learning error rate. Figure 6 shows the results obtained on varying the number of neurons to be used in the input layer. It could be easily analyzed that the neural network has he least error learning rate when the number of neurons to be used in the input layer are 20.

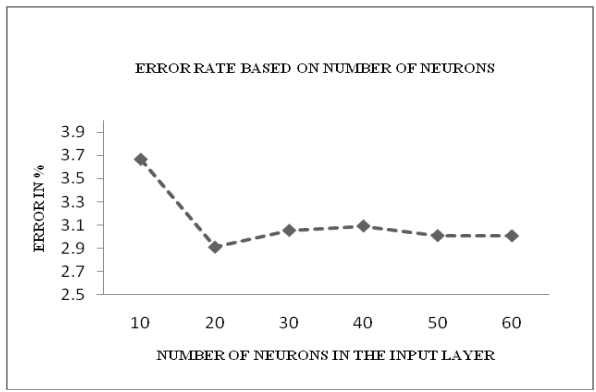

Figure 6: Error Rate Based on Neurons in the Input Layer

We also evaluated the performance of the neural network used in the Relational Behavior Pattern Generator for varying learning rate, number of iterations to be considered. The results obtained are as shown in Fig.7. Based on the results obtained we have understood that the neural network has the least error rate for Learning Rate = 0.2 and the Number of Iterations to be used are fixed to 10000. The response time of the neural network using 20 neurons in input layer, 0.2 is the Learning Rate and with 10000 iterations was approximately 17.249 milliseconds which is acceptable.

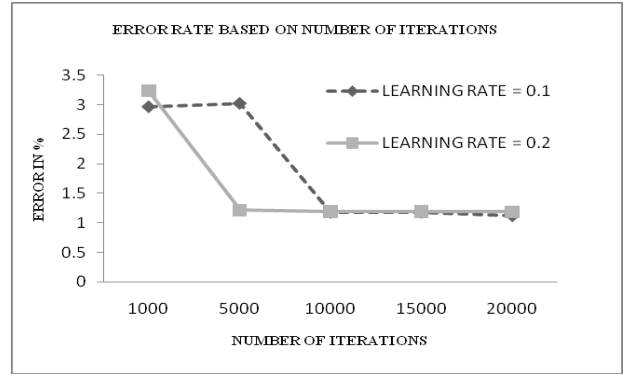

Figure 7: Error Rate Comparisons varying Iterations and Learning Rates

The Policy and Configuration Assignment Block of the CSM also uses the MFNN architecture to detect unauthenticated node activity. The MFNN architecture is also decided based on using similar evaluation represented in Fig.6. The MFNN is constructed using 3 layers having 26000 neurons in the second layer and 8,000 neurons in the third layer. The learning considered is 0.2 and the iterations used are 10000. The MFFN under testing had a response time of about 13.649. It is very clear from the evaluation of both the neural networks used in the CP that the introduction of the intelligence parameters into the network would not affect the network response time by a great extent, which was a factor taken into consideration during design.

After evaluating the intelligence components of the CSM, we have evaluated the performance of the CSM. The Policy and Configuration Management unit of the CSM is responsible for learning about the dynamic network condition of the test WLAN considered. The test node's network activity was deviated from the operational matrix set for it. Similar simultaneous deviations were introduced into the test WLAN. We evaluated the response of the CSM to such kind of attacks and the results obtained are as shown in Fig.8.

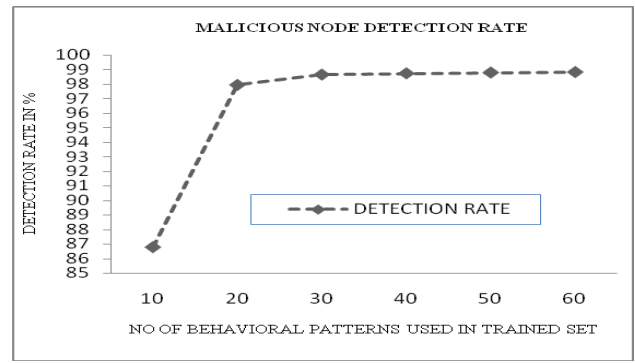

Figure 8: Malicious Node Detection Rate

From Fig.8 it is very clear that as the CSM achieves a very high node misbehavior detection rate. Also it could be seen that the CSM exhibits remarkable learning capabilities. The learning capabilities of the CSM are directly proportional to the training set number provided to the Policy and Configuration Unit. The MFNN used in the Policy and Configuration Unit of the CSM responsible for network learning exhibits remarkable detection rates when well trained.

The experimental evaluation of the CSM performance discussed in this paper was extensive and the CFA proposed exhibits secure cognitive properties.

## V. CONCLUSION

In this paper we analyzed the security threats and deficiencies in our current WLAN implementations. Cognitive Networks have emerged as a new model for designing secure and intelligent WLAN. In our research presented in this paper we proposed the CFA to implement a CP and to secure WLAN's. Through our evaluation of the system it was observed that the CP was achieved using the CSM. Access control mechanism using the PADL also provided highly efficient security mechanism to the cognitive wireless network.

The cognition achieved was highly effective as we used a cross layer adaptive design approach. For the purpose of intelligence implementation, soft computing techniques proved useful.

ACKNOWLEDGMENT

The authors would like to express their cordial thanks to Mr Ashutosh Kumar and Mr Harshad Patel of Planet-*i* Technologies for their much valued support and advice.

AUTHORS PROFILE

**Mr. S.C.Lingareddy** is a PhD Student in Computer Science at Jawaharlal Nehru Technological University Hyderabad. Currently he is working as Assistant Professor and Head of the Department of Computer Science and Engg. KNS Institute of Technology, Bangalore. He received the B.E(CSE). degree from Karnataka University Dharwad and M.Tech.(CSE) degrees from Visvesvaraya Technological University Belgaum. in 1994 and 2004, respectively. He is a member of IEEE, ISTE. CSI, His research interests are Network Security, Information Security, Wireless sensor Network, Cognitive Radio Network.

**Dr. B Stephen Charles** received ME degree from Bharathiar University, Coimbatore and PhD from Jawaharlal Nehru Technological University Hyderabad. He published 18 International journal Papers and 2 National Journal papers,35 International conference papers. He has 23 years of experience in Teaching, He is working as a Principal in Stanley Stephen College of Engineering, Kurnool, His research interests are digital signal processing, Network Security ,Information Security and Wireless network

**Dr. Vinaya Babu** received ME, M.Tech,(CSE) ,PhD degree in Electronics and Communication. Has a total of 30 publications in National and International Journals. Is a member of many professional bodies like IEEE, IETE, ISTE, CSI and was the President and vice-President of Teachers Association, Has a total of 23 years of experience in Teaching. Dr.Vinaya Babu is currently serving as the Director of Admission and Professor of CSE., His area of interests Algorithm, Information Retrieval and Data mining, Computational Models, Computer Networks, Image Processing and Computer Architecture.

**Mr. Kashyap Dhruve** received his Bachelor of Engineering Degree in Electronics and Communication Engineering from Visvesvaraya Technological University Belgaum. He is currently working as a technical director in Planet-i Technologies. His areas of research interests are Information Security, Image Processing, Analog Design of Sensor Interface Circuts , Data Compression, Wireless Networks , Wireless Sensor Networks , Cognitive Networks.